\documentclass[aps,reprint,pra,longbibliography]{revtex4-1}
\usepackage{}
\usepackage{amssymb}
\usepackage{amsmath}
\usepackage{dcolumn}
\usepackage{bm}
\usepackage{graphicx}
\usepackage{mathrsfs}
\usepackage[colorlinks,linkcolor=blue,anchorcolor=blue,citecolor=blue,urlcolor=blue]{hyperref}

\begin{document}

\title{Enhancing the spin-photon coupling with a micromagnet}

\author{Xin-Lei Hei}
\author{Xing-Liang Dong}
\author{Jia-Qiang Chen}
\author{Cai-Peng Shen}
\author{Yi-Fan Qiao}
\author{Peng-Bo Li}
\email{lipengbo@mail.xjtu.edu.cn}
\affiliation{Shaanxi Province Key Laboratory of Quantum Information and Quantum Optoelectronic Devices,\\School of Physics, Xi'an Jiaotong University, Xi'an 710049, China}

\date{\today}

\begin{abstract}
Hybrid quantum systems involving solid-state spins and superconducting microwave cavities play a crucial role in quantum science and technology, but improving the spin-photon coupling at the single quantum level remains challenging  in such systems. Here,
we propose a simple technique to strongly couple a single solid-state spin to the microwave photons in a superconducting coplanar waveguide (CPW) cavity via a magnetic microsphere. We show that, strong coupling  at the single spin level can be realized by virtual magnonic excitations of a nearby micromagnet.
The spin-photon coupling strength can be enhanced up to typically four orders of
magnitude larger than that without the use of the micromagnet. This work can find applications in quantum information processing with strongly coupled solid-state spin-photonic systems.
\end{abstract}

\maketitle

\section{\label{sec:level1} introduction}
Hybrid quantum architectures based on degrees of freedom of completely different nature have attracted much attention over recent years~\cite{RevModPhys.85.623,PhysRevApplied.4.044003,PhysRevLett.118.140501,PhysRevLett.124.113602,Li:20,PhysRevLett.124.053601,PhysRevLett.107.220501,PhysRevLett.109.033604,PhysRevLett.110.156402,PhysRevLett.117.015502,PhysRevApplied.10.024011,PhysRevResearch.2.013121,blais2020quantum,PhysRevResearch.3.013025,Arcizet2011,PhysRevLett.125.153602,PhysRevB.81.241202,PhysRevLett.107.060502,PhysRevB.98.024406,PhysRevLett.110.250503,PhysRevLett.102.083602,PhysRevLett.105.140502}, due to a variety of applications in quantum technologies such as quantum networks \cite{RevModPhys.87.1379}, information processing \cite{RevModPhys.74.347,blais2020quantum} and sensing \cite{RevModPhys.89.035002}.
The interactions between solid-state spins and microwave photons play a central role in hybrid quantum systems \cite{PhysRevLett.105.140502,PhysRevB.81.241202,PhysRevLett.107.060502,PhysRevB.98.024406,PhysRevLett.110.250503,PhysRevLett.102.083602}, which associate the solid-state spins as quantum memories \cite{Fuchs2011} and the photons as quantum information carriers \cite{Zhong1460}.
Besides, they can be useful for fundamental investigation in quantum mechanics, solid-state physics and quantum optics, which provide useful platforms and tools to deepen the research of currently unexploited quantum physics~\cite{Mi2018,Awschalom2018}.

To construct hybrid solid-state platforms, color centers in diamond are often employed.
For example, NV centers, one of the excellent color centers with long coherence time and stable triplet ground states \cite{Aharonovich2011,Lee_2017,DOHERTY20131,PhysRevLett.112.047601,RevModPhys.92.015004,Bar-Gill2013,Abobeih2018One}, are frequently used for quantum storage \cite{Fuchs2011} and sensing~\cite{Dolde2011,Kolkowitz1603}.
While the realization of strong coupling is difficult due to the large mismatch between the spatial extension of free-space photons and typical spins, the solid-state spins can strongly couple to CPW resonators \cite{RevModPhys.89.021001,PhysRevB.81.241202,PhysRevLett.107.060502}, similar to cavity quantum electrodynamics with atoms\cite{PhysRevLett.112.213602}.
However, current experiments in hybrid systems often involve spin ensembles rather than single spins~\cite{PhysRevLett.103.043603,PhysRevLett.105.140502,PhysRevLett.107.060502,PhysRevLett.97.033003}, since the single spin-photon coupling strength is just around 10 Hz, which is far from the strong coupling regime~\cite{Mi2018,blais2020quantum}.

Here, we propose a feasible scheme to realize the strong spin-photon coupling at the single quantum level  via virtual excitations of magnons in a micromagnet. Magnons, the energy quanta of spin waves, play an essential role in quantum information processing and quantum sensing~\cite{Zhang2015, stancil2009spin,Zhange1501286,PhysRevLett.117.133602,Lachance_Quirion_2019,PhysRevLett.125.117701}, with different types of  magnets such as sphere magnets \cite{PhysRevLett.104.077202,PhysRevLett.113.083603,PhysRevLett.113.156401,Tabuchi405,doi:10.1063/1.4907694,PhysRevA.93.021803,PhysRevB.93.144420,doi:10.1063/1.4941730,PhysRevLett.120.057202,PhysRevB.101.125404,PhysRevLett.124.163604}, film layer \cite{PhysRevLett.106.216601,PhysRevLett.111.127003,doi:10.1063/1.4939134,PhysRevLett.123.107701,PhysRevLett.123.107702,PhysRevLett.124.027203,PhysRevLett.125.077203} and cylinder magnets~\cite{Almulhem2020}. Thanks to the small mode volume and high spin density of the Kittel mode in a sphere micromagnet, it can focus the field energy to enable the strong spin-magnon interaction \cite{PhysRevB.101.134415}. Therefore, the Kittel mode in the sphere magnet is frequently used  in the magnon-cavity coupling system to achieve the strong coupling between magnons and microwave cavity photons~\cite{simons2001coplanar,Wolf1488,PhysRevLett.103.043603,PhysRevLett.103.070502,TABUCHI2016729,Morris2017,PhysRevB.95.214423,Lachance.Quirion.2019,PhysRevA.99.021801,acsphotonics.8b00954}.
The nanoscale micromagnet, yttrium iron garnet (YIG) sphere considered in this work, is well matched with the NV spins and CPW cavities, and is very crucial to  the strong interactions among the three subsystems.
We consider efficient coupling of magnonic excitations to a single NV spin and a CPW cavity. We show that, even when the Kittel mode is only virtually populated, the magnon-mediated interaction between the NV spin and the microwave cavity mode can be modified significantly. The induced spin-photon coupling strength can be enhanced up to typically four orders of magnitude larger than that in the absence of the micromagnet. This regime may  provide a powerful tool for applications of quantum information processing based on strong spin-photon interactions at the single quantum level.

\section{\label{sec:level2}description of the system}
\subsection{\label{sec:level2a}The setup}
\begin{figure}[b]
	\includegraphics[width=8.5cm]{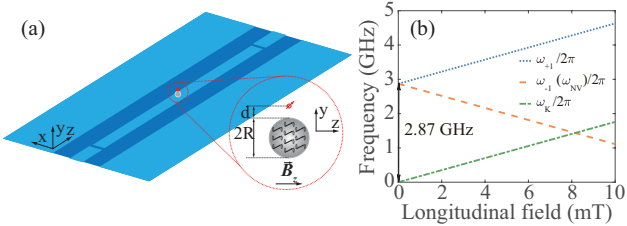}
	\caption{\label{fig:f1}(color online). (a) Schematic of the NV center (red circle with arrow) coupled to a CPW cavity (blue) via a YIG sphere (gray). (b) Frequency split of the NV center spin states and the resonant frequency of the Kittel mode changed with the longitudinal magnetic field.}
\end{figure}

As illustrated in Fig.~\ref{fig:f1}(a), we consider a hybrid tripartite system where a spherical micromagnet with radius $R$ is magnetically coupled to a NV center and a CPW resonator simultaneously.
Here, the magnetic microsphere is a YIG sphere.
The NV center is placed above the microsphere, which is close to the surface of the CPW resonator.

The magnetic microsphere supports spin waves,
which always exist in the ferrimagnetic or antiferromagnetic materials in the low energy limit.
The spin wave in a magnetic microsphere can be quantized by the dipolar, isotropic, and magnetostatic approximations~\cite{PhysRevB.101.125404}, with magnons as the quanta.
In our setup, only the Kittel mode in the YIG sphere is considered, where all the spins in the magnetic sphere precess in phase and with the same amplitude~\cite{TABUCHI2016729}.
The Hamiltonian of the Kittel mode (with the destruction operator $ \hat{s}_K $) can be expressed as
\begin{eqnarray}\label{EQ1}
  \hat{H}_K = \hbar\omega_{K}\hat{s}_K^{\dagger}\hat{s}_K,
\end{eqnarray}
Here, the frequency $ \omega_{K} = |\gamma|B_z $ is controlled by the external magnetic field,
with the gyromagnetic ratio $ \gamma=-1.76\times 10^{11} $ T$ ^{-1} $s$ ^{-1} $.

For the NV center, the spin-triplet $S=1$ ground states \{$ |0\rangle $, $|{\pm 1}\rangle $\} are the eigenstates of the spin operator $ \hat{S}_z $, with $ \hat{S}_z|{i}\rangle = i|{i}\rangle $ and $ i=0,\pm1 $.
In the presence of the external magnetic field $ \vec{B}_z $ oriented along the NV symmetry axis ~\cite{Maze_2011,DOHERTY20131,Lee_2017,PhysRevApplied.11.044026,RevModPhys.92.015004},
the degeneracy of the states $|{\pm 1}\rangle $ are split via the Zeeman effect.
Taking $|{0}\rangle$ as the energy reference,
the Hamiltonian of the NV center can be expressed as
\begin{equation}
 \hat{H}_{NV} = \hbar\sum_{i=\pm1}\omega_{i}|{i}\rangle \langle{i}|,
\end{equation}
 with $ \omega_{\pm1}=D_0\pm|\gamma|B_z $ and the zero-field-splitting $D_0 = 2\pi\times 2.87 $ GHz.
With proper magnetic fields, we can choose the states $|{0}\rangle $ and $|{-1}\rangle $ as a spin qubit, which are selected to be resonant with the magnon mode.
The Hamiltonian of the NV center can be simplified by
\begin{eqnarray}\label{EQ2}
\hat{H}_{NV} = \frac{\hbar}{2}\omega_{NV}\hat{\sigma}_z,
\end{eqnarray}
where we define the frequency as $ \omega_{NV}\equiv \omega_{-1} $ and the spin operator as $ \hat{\sigma_z} \equiv |{-1}\rangle \langle{-1}|-|{0}\rangle \langle{0}|$.
The state $ |{+1}\rangle $ can be safely excluded due to its off-resonance with the Kittel mode.

The CPW resonator is a one dimensional resonator whose frequency $ \omega_{C} $ is related to
the length of the transmission line segment.
The electron oscillating in the line cavity creates a variational electromagnetic field,
which can be quantized and interacts with other devices~\cite{PhysRevLett.103.070502, PhysRevB.84.060501,Jenkins_2013,doi:10.1063/1.4899141,acsphotonics.8b00954}.
The corresponding Hamiltonian of photons in the CPW resonator can be written as
\begin{eqnarray}\label{EQ3}
\hat{H}_{C} = \hbar\omega_{C}\hat{a}^{\dagger}\hat{a},
\end{eqnarray}
where $ \hat{a} $ is the annihilation operator of the photon mode.
The photon frequency is designed to match the frequencies of magnons and spins.
That means, in this setup, the frequencies of the three individual systems are equivalent.
\subsection{\label{sec:level2b}Interactions between spins and magnons}
We now consider the quantization of the magnetic field induced by the magnetic sphere.
This magnetic field exists both inside and outside the micromagnet.
Here, we focus on  the latter because of the location of spin qubit.
From classical electrodynamics, the magnetic field of a magnetic sphere with the magnetization $ \vec{M} $ can be expressed as
\begin{equation}
\vec {B}_m(\vec{r}) = \mu_0 R^3/(3r^3)\{ 3( \vec {M}\cdot\vec{r})\vec{r}/r^2- \vec {M}\},
\end{equation}
where $ \mu_0 = 4\pi\times 10^{-7} $ T$\cdot$m$/$A is the vacuum permeability, $R$ is the radius of the YIG sphere, and $ \vec{r}=(r,\theta,\phi) $ is the position vector relative to the centre of the magnetic sphere.
After quantizing the spin wave, the corresponding magnetization operator $ \hat{\vec{M}} $ of the Kittel mode is
\begin{eqnarray}\label{EQ5}
\hat{ \vec {M}} = {M_{K}} \left( \vec{\tilde{ m}}_K\hat {s}_K + \vec{\tilde{ m}}_K^*\hat {s}_K^{\dagger}\right),
\end{eqnarray}
where $ M_{K} = \sqrt{\hbar |\gamma|M_s/2V} $ is the zero-point magnetization, $ M_s $ is the saturation magnetization, and $ V $ is the volume.
Meanwhile, we define the Kittel mode function as $ \vec{\tilde{m}}_K = \vec{e}_x+i\vec{e}_y $ with the unit coordinate vector $ \vec{e}_x $ and $ \vec{e}_y $ (see more details in Appendix~\ref{app:a}).
Remarkably, there is no $z$ component in the mode function.
From the above discussion, we can get the quantized magnetic field $ \hat{\vec{B}}_m $ as
\begin{eqnarray}\label{EQ6}
\hat{ \vec {B}}_m(\vec{r}) =&& \frac{\mu_0 R^3 M_{K}}{3r^3}\Big\{\big[ (3C_{\theta}^2-1)\hat{X}+3S_{\theta}C_{\theta}\hat{P} \big]\vec{e}_x \nonumber\\
&&+ \big[ 3S_{\theta}C_{\theta}\hat{X} + (3S_{\theta}^2-1)\hat{P} \big]\vec{e}_y \Big\},
\end{eqnarray}
where we define $ C_{\theta}=\cos\theta $, $ S_{\theta}=\sin\theta $, $ \hat{X} = \hat {s}_K + \hat {s}_K^{\dagger} $ and $ \hat{P} = i(\hat {s}_K - \hat {s}_K^{\dagger}) $ for convenience.

\begin{figure}[t]
	\includegraphics[width=8.5cm]{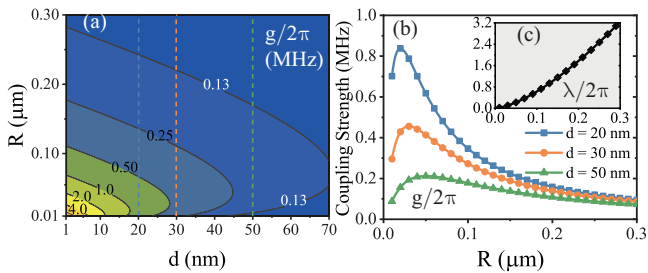}
	\caption{\label{fig:f2}(color online). (a)- (b) The couping strength $ g/2\pi $ between the individual NV center spin qubit and the Kittel mode in the YIG sphere versus the distance $ d $ and the radius $ R $. (c) The coupling strength $ \lambda/2\pi $ between the CPW resonator photon and the Kittel mode in the YIG sphere versus the radius $ R $. }
\end{figure}

We then consider the interaction between the spin qubit and the Kittel mode in the micromagnet.
When a magnetic dipole is placed in a magnetic field, it experiences a torque which tends to line it up parallel to the field.
The Hamiltonian associated with this torque can be naturally written as
\begin{equation}
\hat{H}_{N-K}/\hbar = -g_e\mu_B \hat{ \vec {B}}_m \cdot \hat{\vec{S}}
\end{equation}
with the electronic spin Land\'{e} factor $ g_e $, the Bohr magneton $ \mu_B $ and the spin operator $ \hat{\vec{S}} = (\hat{S}_x,\hat{S}_y,\hat{S}_z) $.
In this work, we define the corresponding components of the spin operator as $ \hat{S}_x = \hbar (|{-1}\rangle\langle{0}| + |{0}\rangle \langle{-1}|)/2 $, $ \hat{S}_y = \hbar (|{-1}\rangle \langle{0}| - |{0}\rangle\langle{-1}|)/2i $.
Assuming that the coupling strength is smaller than the resonance frequency, the spin-magnon interaction under the rotation wave approximation can be described by
\begin{eqnarray}\label{EQ9}
\hat{H}_{N-K}=-\hbar g(\hat{s}_{K} \hat{\sigma}^{+}+H.c.),
\end{eqnarray}
where the spin operators are $ \hat{\sigma}^{+} = |{-1}\rangle \langle{0}| $, and $ \hat{\sigma}^{-} = |{0}\rangle \langle{-1}| $.
And the corresponding spin-magnon coupling strength is expressed as
\begin{eqnarray}\label{EQ8}
g = \sqrt{\frac{|\gamma|M_s}{12\pi}}\frac{g_e \mu_0 \mu_B R^{3/2}}{(R+d)^3},
\end{eqnarray}
where the distance between the spin qubit and the surface of the magnetic sphere as $ d = r-R $ (see Fig.~\ref{fig:f1}(a)).
The coupling strength is proportional to the square root of the micromagnet volume with $ d>R $, while the coupling strength decreases slowly with $ R>d $. In Fig.~\ref{fig:f2}(a) and Fig.~\ref{fig:f2}(b), we show the dependence of the spin-magnon coupling as a function of $ R$ and $d $.
The coupling rate keep reducing with the increase of $ d $. While it increases with the growth of $ R $ utill the maximum.
From the numerical result we find that if we choose the YIG sphere with $ R\sim 50 $ nm and the distance $ d\sim10 $ nm, then the coupling strength is about $ \sim 1 $ MHz.
\subsection{\label{sec:level2c}Interactions between photons and magnons}
We consider the quantizing form of the magnetic field generated by the CPW resonator.
For the single mode, the  magnetic field operator is
\begin{equation}
 \hat{ \vec {B}}_c =\vec{b}_{tr}(x, y)(\hat{a}+ \hat{a}^{\dagger})/\sqrt{2}.
\end{equation}
Here, the mode function $ \vec{b}_{tr}(x,y) $ varies strongly in space depending on the CPW resonator geometry.
To obtain the stronger interaction, the magnetic sphere is in close proximity to the surface of the CPW resonator.
Then the mode function depends on the radius of the sphere ($ \vec{b}_{tr}(R) $).
From the discussion in Sec.\ref{sec:level2a}, the radius is about tens of nanometers, which is less than the width of the CPW resonator ($ \sim 1$ $\mu $m).
Therefore, the variation of the mode function depending on the radius can be ignored and $ \vec{b}_{tr}(R) $ acts as an estimated constant ranging from $ 35\sim40 $ $\mu$G.

We then expound the magnetic interaction between the CPW resonator and the micromagnet.
The classical exchange energy of a magnetic dipole with the magnetic moment $ \vec{m} $ in a uniform magnetic field $ \vec{B}_c $ is $ U = -\vec{m}\cdot\vec{B}_c $.
For a uniformly magnetized sphere, we treat it as a magnetic dipole with $ \vec{m}=V\vec{M} $.
After replacing the classical quantity with the magnetization operator $\hat{ \vec {M}}$ and the magnetic field operator $ \hat{ \vec {B}}_c $ in the expression above, the photon-magnon interaction Hamiltonian is
\begin{equation*}
\hat{H}_{C-K} = -\hbar \lambda(\hat {s}_K + \hat {s}_K^{\dagger}) (\hat{a}+ \hat{a}^{\dagger}).
\end{equation*}
Here, the homologous coupling strength is defined as
\begin{eqnarray}\label{EQ11}
\lambda=\sqrt{\frac{\pi |\gamma|M_s }{3\hbar}}b_x(R)R^{3/2}.
\end{eqnarray}
We ignore the $ y $ component of $\vec{b}_{tr}(R) $, since the magnetic field is parallel to the $ x $ axis.
Then, the photon-magnon coupling strength $ \lambda $ can be enhanced to around $ 1 $ MHz with a proper geometrical size of the magnetic microsphere (see Fig.~\ref{fig:f2}(c)).
Under the rotating wave approximation, we can get the photon-magnon interaction Hamiltonian
\begin{eqnarray}\label{EQ12}
\hat{H}_{C-K}= -\hbar \lambda( \hat {s}_K^{\dagger}\hat{a} + H.c. ).
\end{eqnarray}

\section{\label{sec:level3}realistic consideration and experimental parameters}
From the above discussion, we can obtain the total Hamiltonian of the hybrid spin-magnon-photon system, which can be expressed as
\begin{eqnarray}\label{EQ13}
\hat{H} =&& \hbar\omega_C \hat{a}^{\dagger}\hat{a}+\hbar\omega_K \hat{s}_K^{\dagger}\hat{s}_K+\frac{1}{2}\hbar\omega_{NV} \hat{\sigma}_{z}\nonumber\\
&&-\hbar g(\hat{s}_{K} \hat{\sigma}^{+}+H.c.)- \hbar \lambda( \hat {s}_K^{\dagger}\hat{a} + H.c. ),
\end{eqnarray}
where the first three terms correspond to the free Hamiltonian from Eq.~(\ref{EQ1}), Eq.~(\ref{EQ2}) and Eq.~(\ref{EQ3}), and the last two terms describe two interactions: one between the Kittel mode and the  NV center spin  from Eq.~(\ref{EQ9}), while the other between the Kittel mode and the photon in the CPW resonator from Eq.~(\ref{EQ12}).
As the Kittel mode is coupled both to the  spin qubit and the CPW resonator photon, it is possible to work as a quantum interface between the spin qubit and the CPW resonator.

We now discuss the  dynamics of the system in a realistic
situation.
In this case, we take into account  the dephasing of the  NV center spin
($ \gamma_s $), the decays of the Kittel mode ($ \gamma_m $) and the CPW resonator ($ \kappa $).
As a result, the  master equation of the total system can be expressed as
\begin{equation}
   {\dot{\hat \rho} (t)} =  -i/ \hbar[ {\hat H,\hat \rho } ] + \mathcal{L}[\hat{\rho}],
\end{equation}
where $\hat \rho $ is the density operator. The last term in above equation is given by
\begin{eqnarray}
 \mathcal{L}[\hat{\rho}] &=& {\gamma _s}\mathcal{D}[ {\hat \sigma _z} ]\hat \rho    + \sum_{j}\{(\bar{n}_{j} + 1){\Gamma _j}\mathcal{D}[ {{{\hat o}_j}} ]\hat \rho\nonumber\\&& +\bar{n}_{j}{\Gamma _j}\mathcal{D}[ \hat {o}_j^{\dagger} ]\hat \rho\}.
\end{eqnarray}
For compactness we define $ \{ \Gamma_m, \Gamma_p \}\equiv\{ \gamma_m, \kappa \} $, $ \{\hat{o}_m,\hat{o}_p\} \equiv \{\hat{s}_K,\hat{a}\} $ and $ \mathcal{D}[\hat{o}]\hat{\rho} \equiv \hat{o}\hat{\rho}\hat{\rho}^\dagger -\{\hat{o}^\dagger\hat{o}, \hat{\rho}\}/2 $.
Here, we have introduced the thermal occupation number $ \bar{n}_j = (e^{\hbar\omega_j/kT} - 1)^{-1} $ with frequencies $ \{\omega_m,\omega_p\}=\{\omega_{K},\omega_{C}\} $ and environmental temperature $ T $~\cite{PhysRevApplied.4.044003,PhysRevLett.124.093602}.
Considering the low temperature limit ($ T\sim 10 $ mT), the thermal occupation number $ \bar{n}_j $ is safely ingnored with the frequency $ \omega_{j}\sim 1.4 $ GHz.
Then the dissipation of the Kittel mode and the photon mode therefore can be simplified.
Then, the master equation turns to
\begin{equation}
   {\dot{\hat \rho} (t)} =  -i/ \hbar[ {\hat H,\hat \rho } ] + {\gamma _s}\mathcal{D}[ {\hat \sigma _z} ]\hat \rho    + \sum_{j}{\Gamma _j}\mathcal{D}[ {{{\hat o}_j}} ]\hat \rho.
\end{equation}

In the frame rotating at the individual spin qubit frequency $ \omega_{NV} $, the whole Hamiltonian is firstly transformed to the form as
\begin{eqnarray}
\hat{H} &=& \hbar\Delta_1 \hat{s}_K^{\dagger}\hat{s}_K+\hbar\Delta_2 \hat{a}^{\dagger}\hat{a}-\hbar g\hat{s}_{K} \hat{\sigma}^{+}\\ \nonumber
&&- \hbar \lambda\hat {s}_K^{\dagger}\hat{a} + H.c.,
\end{eqnarray}
where $ \Delta_1=\omega_{K}-\omega_{NV} $ and $ \Delta_2=\omega_{C}-\omega_{NV} $~\cite{PhysRevLett.112.213602}.
In view of the large detuning ($ \Delta_1\gg g,$ $\lambda $), the Kittel mode can be eliminated~\cite{PhysRevA.85.042306} and we can obtain the effective interaction between the NV center spin qubit and the CPW resonator photon with the following Hamiltonian
\begin{eqnarray}\label{EQ16}
\hat{H}_{\mathrm{eff}} = &&\hbar(\Delta_2-\beta^2\Delta_1)\hat{a}^{\dagger}\hat{a}-\frac{1}{2}\hbar\alpha^2\Delta_1 \hat{\sigma}_{z} \nonumber\\
&&-\hbar g_{\mathrm{eff}}(\hat{a}^{\dagger}\hat{\sigma}^{-} + H.c.),
\end{eqnarray}
where we difined the dimensionless parameters as $ \alpha = g/\Delta_1 $, $ \beta = \lambda/\Delta_1 $ as well as the effective coupling srength $ g_{\mathrm{eff}} = g\lambda/\Delta_1 $.
This result shows that the interface between  the spin and  the cavity photon is achieved via utilizing the Kittel mode in the micromagnet as a medium.

We next discuss the effective coupled system.
First, the dissipative interaction system can be discribed with a reduced effective master equation as
\begin{eqnarray}\label{EQ17}
\frac{{d{{\hat \rho }_r}(t)}}{{dt}} =  &&- \frac{i}{\hbar }[ {{{\hat H}_{\mathrm{eff}}},{{\hat \rho }_r}} ] + {\gamma _s}\mathcal{D}[ {\hat \sigma_z} ]{{\hat \rho }_r} \nonumber\\
&&+ {\gamma _{\mathrm{eff}}}\mathcal{D}[ {{{\hat \sigma }_- }} ]{{\hat \rho }_r} + {\kappa _{\mathrm{eff}}}\mathcal{D}[ {\hat a} ]{{\hat \rho }_r},
\end{eqnarray}
where  $ {\gamma _{\mathrm{eff}}} =\ {\alpha ^2}{\gamma _m} $ and $ {\kappa _{\mathrm{eff}}} = \kappa  + {\beta ^2}{\gamma _m} $ are the effective dacay rates, respectively.
Here, the original dissipation rates are chosen as $ \gamma_s/2\pi\sim 1$ kHz~\cite{PhysRevApplied.11.044026}, $\gamma_m/2\pi\sim 1$ MHz~\cite{PhysRevLett.124.093602} and $\kappa/2\pi\sim 6 $ kHz~\cite{PhysRevApplied.4.044003}.
And we take the dimensionless parameters as $ \alpha \sim \beta \sim 0.1 $ for the condition of eliminating the Kittel mode.
We therefore estimate the effective dacay rates as $ {\gamma _{\mathrm{eff}}} \sim 10 $ kHz and $ {\kappa _{\mathrm{eff}}} \sim 16 $ kHz.
At the same time, the effective coupling strength $ g_{\mathrm{eff}} $ can be approximatively obtained as $ g_{\mathrm{eff}}\sim 0.1g $, which is larger than the effective dacay rates $ {\gamma _{\mathrm{eff}}} $, $ {\kappa _{\mathrm{eff}}} $ as well as the dephasing of the NV center $ \gamma_s $. Since the spin dephasing rate $ \gamma_s $ is much smaller than the effective decay rates $ {\gamma _{\mathrm{eff}}}$ and ${\kappa _{\mathrm{eff}}}$, its effect on the dynamics of the system can be neglected.
Therefore, we can obtain the center result of this work: the individual NV center spin qubit can strongly couple to the microwave photon in  a CPW resonator  under the proper conditions. The spin-photon coupling is well within the strong coupling regime.
\begin{figure}[t]
	\includegraphics[width=8.5cm]{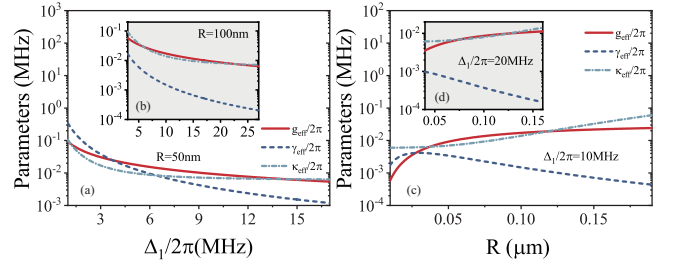}
	\caption{\label{fig:f3}(color online). The effective coupling strength $ g_{\mathrm{eff}}/2\pi $, the effective decay rates $ {\gamma _{\mathrm{eff}}/2\pi} $ and $ {\kappa _{\mathrm{eff}}}/2\pi $ depend on the detuning $ \Delta_1/2\pi $ with the value of (a) $ R=50 $ nm and (b) $ R=100 $ nm. And inversely these parameters depend on the radius of the micromagnet sphere $ R $ with the value of (c) $ \Delta_1/2\pi=10 $ MHz and (d) $ \Delta_1=20 $ MHz. We  choose the distance  between the spin qubit and the surface of the magnetic microsphere as $ d=30 $ nm in these figures. }
\end{figure}
Accurately, the effective coupling strength and decay rates depend on the detuning $ \Delta_1 $ and the geometrical radius of the magnetic sphere $ R $, which is clearly shown in Fig.~\ref{fig:f3}. The effective coupling rate $ g_{\mathrm{eff}}/2\pi $ and decay rate $ {\kappa _{\mathrm{eff}}}/2\pi $ increase with the increase of the detuning $ \Delta_1/2\pi $ while they decrease with the increase of $ R $. Another effective decay rate $ {\gamma _{\mathrm{eff}}/2\pi} $ decreases with the increase of $ \Delta_1/2\pi $ and $ R $. There is always a large range where the effective coupling strength exceeds both the effective decay rates with various values of detuning $ \Delta_1/2\pi $ and $ R $. Thus, we can take the effective spin-photon  system into the strong coupling regime via tuning the relevant parameters.

\begin{figure}[t]
	\includegraphics[width=8.5cm]{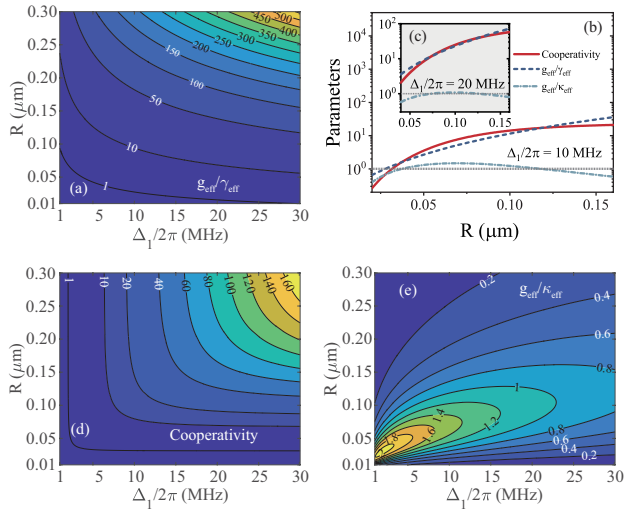}
	\caption{\label{fig:f4}(color online). The contour maps show the parameters (a) $ g_{\mathrm{eff}}/\gamma _{\mathrm{eff}} $, (d) cooperativity and (e) $ g_{\mathrm{eff}}/\kappa _{\mathrm{eff}} $ as functions of the detuning $ \Delta_1/2\pi $ and the radius $ R $. The more visualized relationship among the three parameters, the detuning and the radius is shown in (b) and (c). And the distance between the spin qubit and the surface of the magnetic microsphere is chosen as $ d=30 $ nm as well. The short dots horizontal lines in (b) and (c) point to the value 1.}
\end{figure}

We proceed to analyze the effective spin-photon coupling  from another point of view.
In order to compare the effective coupling strength with the effective dacay rates, we employ three parameters  $ g_{\mathrm{eff}}/\gamma _{\mathrm{eff}} $, $ g_{\mathrm{eff}}/\kappa _{\mathrm{eff}} $ as well as the cooperativity $ {\cal C} = g_{\mathrm{eff}}^2/\gamma _{\mathrm{eff}}\kappa _{\mathrm{eff}} $.
The patameter $ g_{\mathrm{eff}}/\gamma _{\mathrm{eff}} $ increases with the increase of the detuning $ \Delta_1 $ and the radius $ R $ as shown in the Fig.~\ref{fig:f4}(a). The results show that the larger physical dimension of the magnetic microsphere allows for stronger effective interaction between the spin qubit and the CPW resonator photon, which makes experimental realization handy.
There is a  range for the parameter $ g_{\mathrm{eff}}/\kappa _{\mathrm{eff}}>1 $ as shown in the Fig.~\ref{fig:f4}(e).
The cooperativity $ \cal C $ increases with the increase of the detuning $ \Delta_1 $ and the radius $ R $.
And there is a very large range where the cooperativity even exceeds 10 as shown in Fig.~\ref{fig:f4}(d).
Figs.~\ref{fig:f4}(b) and (c) respectively show that the three parameters are larger than 1 with different detunings  when the radius $ R $ is within certain limits.
Furthermore, the results displayed in  Fig.~\ref{fig:f3} and Fig.~\ref{fig:f4} are obtained when the NV spin qubit is close to the magnetic microsphere ($ d=30 $ nm).

We now explain the significance of the magnetic microsphere in this spin-photon coupled system using the numerical simulation.
The direct interaction between the individual spin qubit and the single photon is very weak on account of the low magnetic effect generated by a single microwave photon.
Nevertheless, the magnon mode  is sensitive to weak magnetic fields.
At the same time,  individual spin qubits can strongly couple to the magnon mode with a small distance apart from the magnetic microsphere.
Hence, the magnons play a pivotal role in the realization of strong spin-photon coupling.
The effective interaction can be proportionally enhanced with the increase of the volume of the magnetic microsphere within a certain range.

\begin{figure}[t]
	\includegraphics[width=7cm]{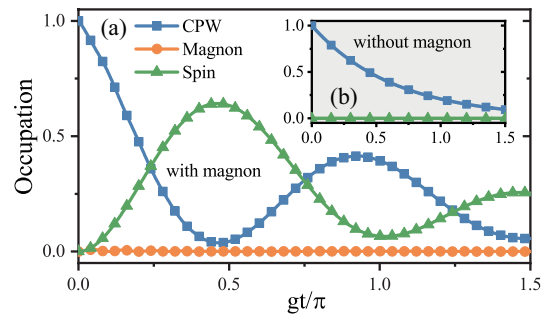}
	\caption{\label{fig:f5}(color online). (a) The occupation map of the three parts in system depends on the dimensionless time $ gt/\pi $ with the parameters $ \gamma_m\sim g $, $ \gamma_s\sim 0.1g $, $ \kappa\sim 0.5g $, $ g\sim\lambda $ and $ \Delta_1\sim 10g $. And the figure (b) shows the direct interaction between the spin qubit and the CPW resonator photon without the magnon mode as an intermediary. }
\end{figure}

We numerically simulate the time evolution of the occupations of the spin qubit, the Kittel magnon and the CPW resonator  using the qutip package in python~\cite{JOHANSSON20131234} as shown in the Fig.~\ref{fig:f5}.
The damped oscillations of the occupations of the spin and photon in  Fig.~\ref{fig:f5}(a) show the strong coupling between them in the presence of magnons.
While in the absence of magnons, Fig.~\ref{fig:f5}(b) shows an exponential decay curve without oscillations.
This result confirms the validity of the effective master equation (\ref{EQ17}).
Note that the population of the virtually excited Kittel mode is nearly zero in the whole process.
These results clearly show the significant role played by the magnon in realizing the strong coupling between the spin and the photon.

\section{\label{sec:level4}Conclusion}
In conclusion, we have proposed  a scheme for strongly coupling a  single solid-state spin like NV centers to the microwave photons in a CPW cavity via  a magnetic microsphere. We have shown that the strong coupling at the single quantum level can be realized by virtual magnonic excitations of a nearby micromagnet.
In contrast to the case in the absence of the micromagnet, the spin-photon coupling strength has been enhanced up to typically four orders of magnitude. Here, the employment of magnons opens up intriguing perspectives for magnonics and spintronics as well. This regime may facilitate much more powerful applications in quantum information processing based on strongly coupled solid-state spin-photonic systems.

\begin{acknowledgments}
This work was supported by the National Natural Science
Foundation of China under Grants No. 92065105, No.
11774285 and the Natural Science Basic Research Program
of Shaanxi (Program No. 2020JC-02).
\end{acknowledgments}

\appendix

\section{\label{app:a}Quantization of the spin wave}

In this Appendix, we integrally show the quantization of the spin wave and obtain the theoretical model of the magnon in a ferromagnetic microsphere.
First, in Sec. \ref{app:a1}, we introduce the original equation of motion for the magnetization, the general Landau-Lifshitz equation, which is simplified under several physical approximations.
Then, we present the Walker mode in a ferrite sphere in Sec.~\ref{app:a2}.
Based on the magnetostatic energy density expression,  we obtain the intrinsic Hamiltonian and quantization of the magnon modes in Sec.~\ref{app:a3}.

\subsection{\label{app:a1}Spin wave equations in a magnetic sphere}
First, we consider the spin waves with a continuous magnetization field $ \vec{M}(\vec{r},t) $ and the corresponding eletromagnetic field intensity $ \vec{E}(\vec{r},t) $ and $ \vec{H}(\vec{r},t) $.
The dynamics of spin waves generally follows the Maxwell's equations with the relationship between the induced magnetization and the applied field.
Now we start from the phenomenological Landau-Lifshitz equation of motion for the magnetization~\cite{stancil2009spin,aharoni2000introduction}
\begin{equation}\label{a1}
\frac{d}{d t} \vec{M}(\vec{r}, t)=-|\gamma| \mu_{0} \vec{M}(\vec{r}, t) \times \vec{H}_{\mathrm{eff}}(\vec{M}, \vec{r}, t),
\end{equation}
where the effective field $ \vec{H}_{\mathrm{eff}}(\vec{M}, \vec{r}, t) $ comprises the Maxwellian field $ \vec{H}(\vec{r},t) $ and the extra parts $ \vec{H}'(\vec{M}, \vec{r}, t) $ as follows~\cite{stancil2009spin}
\begin{equation}\label{a2}
\vec{H}'(\vec{M}, \vec{r}, t)= \vec{H}_{\mathrm{ex}}(\vec{M},\vec{r},t) + \vec{H}_{\mathrm{an}}(\vec{M},\vec{r},t) + \vec{H}_{\mathrm{dm}}(\vec{M},\vec{r},t),
\end{equation}
Here, $ \vec{H}_{\mathrm{ex}} $, $ \vec{H}_{\mathrm{an}} $ and $ \vec{H}_{\mathrm{dm}} $ are effective fields due to exchange, anisotropy and demagnetization induced by the magnetic dipole-dipole interactions, respectively.
The extra part of the effective field evidently depends on the magnetization, which means that the Landau-Lifshitz equation is inhomogeneous.

We then assume that the magnet is in the saturated magnetic state along the $ z $ axis under the collinear field.
Then the fluctuation of the magnetization and the field is very small compared with the constant part~\cite{PhysRevB.101.125404}.
Naturally, they are written as
\begin{subequations}
	\begin{eqnarray}
	\vec{M}(\vec{r},t)&&=M_{S}\vec{e}_z+\vec{m}(\vec{r},t),\label{a3a}\\
	\vec{H}(\vec{r},t)&&=H_{0}\vec{e}_z+\vec{h}(\vec{r},t)\label{a3b}
	\end{eqnarray}
\end{subequations}
Here, $ \vec{m}\ll M_S $ and $ \vec{h}\ll H_0 $ are the dynamical variables to be solved and $ \vec{e}_z $ is the unit vector along the $z$ axis.
Then we consider the extra terms in the effective field.
The exchange field is associated with the domain wall interaction, which is less than the dipole-dipole interaction when the micromagnet sizes is large compared with the domain wall length.
The magnetocrystalline anisotropy for a cubic material is also negligible~\cite{aharoni2000introduction,stancil2009spin,PhysRevB.95.214423}.
As for the demagnetizing field, we assume $ \vec{H}_{\mathrm{dm}}=-(M_S/3)\vec{e}_z $, which is suitable for a spherical magnet~\cite{aharoni2000introduction,jackson2007classical}.
The above expressions conbined with the Eq.~(\ref{a1}) lead to a more clear form of Landau–Lifshitz equation as
\begin{eqnarray}\label{a4}
	\frac{\dot{\vec{m}}(\vec{r}, t)}{|\gamma| \mu_{0} M_{S} H_{0}}&&-\vec{e}_{z} \times\left[\frac{\vec{m}(\vec{r}, t)}{M_{S}}-\frac{\vec{h}(\vec{r}, t)}{H_{0}}\right]+\left[\vec{e}_{z}+\frac{\vec{m}(\vec{r}, t)}{M_{S}}\right]\nonumber\\
	&&\times \frac{(H_0-M_S/3)\vec{e}_z}{H_{0}}=\frac{\vec{h}(\vec{r}, t)}{H_{0}} \times \frac{\vec{m}(\vec{r}, t)}{M_{S}}
\end{eqnarray}
Note that the variables $ \vec{m}/M_S $ and $ \vec{h}/H_0 $ are small.
Thus, we can safely neglect the second order term in the above equation.
Then the Landau-Lifshitz equations is linearized as~\cite{stancil2009spin}
\begin{equation}\label{a5}
	\left[\begin{array}{c}
	\dot{m}_{x}(\vec{r}, t) \\
	\dot{m}_{y}(\vec{r}, t)
	\end{array}\right]=\left[\begin{array}{c}
	-\omega_{0} m_{y}(\vec{r}, t)+\omega_{M} h_{y}(\vec{r}, t) \\
	\omega_{0} m_{x}(\vec{r}, t)-\omega_{M} h_{x}(\vec{r}, t)
	\end{array}\right],
\end{equation}
where,  two relevant system frequencies are
\begin{subequations}
\begin{eqnarray}
	\omega_{\mathrm{M}} &&= |\gamma| \mu_{0} M_{\mathrm{S}},\label{a6a}\\
	\omega_{0} &&= |\gamma| \mu_{0} (H_{0}-M_S/3).\label{a6b}
\end{eqnarray}
\end{subequations}
We notice that the $ z $ component vanishes because of the cross product with vector $ \vec{e}_z $.
The result is not abnormal when we consider only small fluctuations  around the fully magnetized state which is along $ z $ axis.

Finally, it comes to the magnetostatic approximation $ \nabla \times \vec{h}(\vec{r},t)\simeq 0 $.
In the Maxwell equations, the  electric field of the spin wave is uncoupled from $ \vec{h} $ under this circumstance.
In addition, the approximation makes it easy to introduce the magnetostatic potential through $ \vec{h}(\vec{r},t)=-\nabla\psi(\vec{r},t) $.
Combined with the zero-divergence condition $ \nabla\cdot \vec{b} = 0 $ in Maxwell equations, the relation $ \vec{b} = \mu_{0}(\vec{h}+\vec{m}) $ allows one to obtain the following equation for three acalar fileds~\cite{stancil2009spin}
\begin{equation}\label{a7}
	\nabla^{2} \psi(\vec{r}, t)=\partial_{x} m_{x}(\vec{r}, t)+\partial_{y} m_{y}(\vec{r}, t),
\end{equation}
which is the constraint inside the micromagnet.
While the equation outside the micromagnet is $ \nabla^2\psi=0 $.
Therefore, the linear scalar equations Eq.~(\ref{a5}) and Eq.~(\ref{a7}) completely describe the spin wave with the boundary conditions, which is the continuity of the normal direction components of $ \vec{h} $ and $ \vec{b} $, respectively.
The spin-wave eigenmodes solved from these equations are the magnetostatic dipolar spin waves or Walker modes~\cite{aharoni2000introduction,stancil2009spin}.

\subsection{\label{app:a2}Walker modes and Kittel modes}
This section reveals the process of calculating the Walker modes.
First, the expressions of the magnetization and magnatic fields in  terms of the eigenmodes is shown by
\begin{subequations}
	\begin{eqnarray}
	\vec{m}(\vec{r}, t)=\sum_{\beta}\left[s_{\beta} \vec{m}_{\beta}(\vec{r}) e^{-i \omega_{\beta} t}+\mathrm{c.c.}\right],\label{a8a}\\
	\vec{h}(\vec{r}, t)=\sum_{\beta}\left[s_{\beta} \vec{h}_{\beta}(\vec{r}) e^{-i \omega_{\beta} t}+\mathrm{c.c.}\right].\label{a8b}
	\end{eqnarray}
\end{subequations}
Here, the eigenmode fields $ \vec{m}_{\beta}(\vec{r}) $ and $ \vec{h}_{\beta}(\vec{r})=-\nabla\psi_{\beta}(\vec{r}) $ are characterized by a series of mode indices $ \{\beta\} $, an eigenfrequency $ \omega_{\beta} $ and a complex amplitude $ s_{\beta} $~\cite{doi:10.1063/1.1735216,doi:10.1002/pssb.2220820102}.
Then the linearized Landau-Lifshitz equations turns to time-independent forms
\begin{subequations}
	\begin{eqnarray}
	i \omega m_{x}(\vec{r})=\omega_{M} \partial_{y} \psi(\vec{r})+\omega_{0} m_{y}(\vec{r}),\label{a9a}\\
	i \omega m_{y}(\vec{r})=-\omega_{M} \partial_{x} \psi(\vec{r})-\omega_{0} m_{x}(\vec{r}).\label{a9b}
	\end{eqnarray}
\end{subequations}
We can eliminate the scalar field $ m_x(\vec{r}) $ and $ m_y(\vec{r}) $ through these equations and Eq.~(\ref{a7}).
Then the formula only contains the magnetostatic potential as
\begin{subequations}
\begin{equation}\label{a10a}
	\nabla^{2} \psi_{\text {out }}(\vec{r})=0,
\end{equation}
\begin{equation}\label{a10b}
	\left(1+\chi_{p}\right)\left(\frac{\partial^{2}}{\partial x^{2}}+\frac{\partial^{2}}{\partial y^{2}}\right) \psi_{\mathrm{in}}(\vec{r})+\frac{\partial^{2}}{\partial z^{2}} \psi_{\mathrm{in}}(\vec{r})=0,
\end{equation}
\end{subequations}
where $ \psi_{\text {in }} $ and $ \psi_{\text {out }} $ are the magnetostatic potential inside and outside the micromagnet, respectively.
Here, the diagonal element of the Polder susceptibility tensor is defined as~\cite{stancil2009spin}
\begin{equation}\label{a11}
	\chi_{p}(\omega) \equiv \frac{\omega_{M} \omega_{0}}{\omega_{0}^{2}-\omega^{2}}
\end{equation}

As for the outside situation, the general solution in the spherical coordinates is given by
\begin{equation}\label{a12}
	\psi_{\text {out }}(\vec{r})=\sum_{l m}\left[\frac{A_{l m}}{r^{l+1}}+B_{l m} r^{l}\right] Y_{l}^{m}(\theta, \phi).
\end{equation}
Here, the expansion coefficients $A_{l m}$, $ B_{l m} $ are determined by the boundary conditions. The spherical harmonics $ Y_{l}^{m}(\theta, \phi) $ indicates the symmetry of the spin wave mode.
While the potential inside the sphere is more hard to solve.
A set of nonorthogonal coordinates $ \{\xi,\eta,\phi\} $ is introduced to make it convenient.
They fulfill
\begin{subequations}
	\begin{eqnarray}
	x&&=\sqrt{\chi_{p}} R \sqrt{\xi^{2}-1} \sin \eta \cos \phi, \label{a13a}\\
	y&&=\sqrt{\chi_{p}} R \sqrt{\xi^{2}-1} \sin \eta \sin \phi, \label{a13b}\\
	z&&=\sqrt{\frac{\chi_{p}}{1+\chi_{p}}} R \xi \cos \eta. \label{a13c}
	\end{eqnarray}
\end{subequations}
In the above coordinates, the solution of Eq.~(\ref{a10b}) becomes available, which is an expression including Legendre polynomials and spherical harmonics~\cite{doi:10.1063/1.1735216,doi:10.1002/pssb.2220820102} as
\begin{equation}\label{a14}
	\psi_{\mathrm{in}}(\vec{r})=\sum_{l m} C_{l m} P_{l}^{m}(\xi) Y_{l}^{m}(\eta, \phi).
\end{equation}
In each term of the summation, the coefficients $ C_{lm} $ depend on the boundary conditions as well.

We then determine all the coefficients by the boundary conditions.
First, the term corresponding to $ B_{lm} $ is not convergent at infinity, which should be removed to make the potential $ \psi $ regular.
That indicates the first condition, $ B_{lm}=0 $.
Considering the potential on the surface of the sphere, the coordinates are
\begin{equation}\label{a15}
	\xi \rightarrow \xi_{0}=\sqrt{\frac{1+\chi_{p}}{\chi_{p}}}, \quad\{\eta, \phi\} \rightarrow\{\theta, \phi\}.
\end{equation}
Applying these coordinates to the two solution expressions Eq.~(\ref{a12}) and Eq.~(\ref{a14}), we can use the boundary condition, the normal-direction-component continuity of $ \vec{h} $, and obtain the second condition
\begin{equation}\label{a16}
	A_{l m}=C_{l m} P_{l}^{m}\left(\xi_{0}\right) R^{l+1}.
\end{equation}
Similarly, we can apply the  continuity of the normal component of the $\vec{b}$ field to obtain the final condition~\cite{doi:10.1063/1.1735216}
\begin{equation}
	\left.\frac{\partial \psi_{\text {out }}}{\partial r}\right|_{r=R}=\left.\frac{\xi_{0}}{R} \frac{\partial \psi_{\text {in }}}{\partial \xi}\right|_{r=R}-\left.i \frac{\kappa_{p}}{R} \frac{\partial \psi_{\text {in }}}{\partial \phi}\right|_{r=R}.
\end{equation}
Here, $ \kappa_{p}(\omega) = \omega_{M}\omega/(\omega_{0}^2-\omega^2) $ is the off-diagonal element of the Polder susceptibility tensor~\cite{stancil2009spin}.
Applying the above fomula and Eq.~(\ref{a16}), the Walker mode eigenfrequency fulfill the equation as~\cite{doi:10.1063/1.1735216,doi:10.1002/pssb.2220820102}
\begin{equation}
	\xi_{0}(\omega) \frac{P_{l}^{\prime m}\left(\xi_{0}(\omega)\right)}{P_{l}^{m}\left(\xi_{0}(\omega)\right)}+m \kappa_{p}(\omega)+l+1=0.
\end{equation}
There are two results we can obtain from this equation: a) the eigenfrequency $ \omega $ is independent of the radius $ R $; b) there are several solutions for some pairs $ \{l,m\} $ which are not positive and physical ($ l=0 $, for instance).
Generally, we can use the indices $ \{l,m\} $ to mark the allowed eigenmodes of the spin waves.
The  corresponding mode functions have been explicitly calculated in Ref.~\cite{doi:10.1063/1.1735216}.
\subsection{\label{app:a3}The intrinsic Hamiltonian and quantization of the magnetostatic dipolar magnon modes}
We now show the quantization of the Walker modes from a phenomenological micromagnetic energy functional~\cite{PhysRevB.101.125404}
\begin{equation}
	E_{m}(\{\vec{m}\},\{\vec{h}\})=\frac{\mu_{0}}{2} \int d V \vec{m}(\vec{r}, t) \cdot\left[\frac{H_{I}}{M_{S}} \vec{m}(\vec{r}, t)-\vec{h}(\vec{r}, t)\right]
\end{equation}
For convenience, we apply the linearized Landau-Lifshitz equations~(\ref{a5}) to the above expression and the energy only depends on $ \vec{m}(\vec{r},t) $ in the form
\begin{eqnarray}\label{a20}
	E_{m}(\{\vec{m}\})=&&\frac{1}{2|\gamma| M_{S}} \int d V\left(m_{x}(\vec{r}, t) \frac{\partial m_{y}(\vec{r}, t)}{\partial t}\right.\nonumber\\
	&&\left.-m_{y}(\vec{r}, t) \frac{\partial m_{x}(\vec{r}, t)}{\partial t}\right).
\end{eqnarray}
The key of quantization is to find the proper zero-point fluctuation and the analogue energy expression with the harmonic oscillator.
Thus, making use of the magnetization expanded in terms of the eigenmodes, Eq.~(\ref{a8a}), we transform the above energy expression to~\cite{MILLS200616,PhysRev.105.390}
\begin{equation}\label{a21}
	E_{m}=\frac{1}{2\hbar|\gamma| M_{S} } \sum_{\beta} \hbar \omega_{\beta} \Lambda_{\beta}\left[s_{\beta} s_{\beta}^{*}+s_{\beta}^{*} s_{\beta}\right],
\end{equation}
where the coefficients have the following form
\begin{equation}\label{a22}
	\Lambda_{\beta} = 2 \mathrm{Im} \int d V m_{\beta y}(\vec{r}) m_{\beta x}^{*}(\vec{r}).
\end{equation}
Compared with the Hamiltonian of the harmonic oscillator, we can choose adequate eigenmode normalization to fulfill $ \Lambda_{\beta}/(M_S|\gamma|\hbar)=1 $ and the energy expression becomes $ E_m = \hbar/2\sum_{\beta}\omega_{\beta}[s_{\beta} s_{\beta}^{*}+s_{\beta}^{*} s_{\beta}] $.
After replacing the expansion coefficients with the bosonic magnon operators, i.e., $ s_{\beta} \rightarrow \hat{s}_{\beta} $ and $ s_{\beta}^{*} \rightarrow \hat{s}_{\beta}^{\dagger} $ with the commutation relation $ [\hat{s}_{\beta}, \hat{s}_{\beta}^{\dagger}] = 1 $, we can obtain the quantized magnon modes Hamiltonian $ \hat{H}_{m} = \sum_{\beta} \hbar\omega_{\beta} [\hat{s}_{\beta}^{\dagger} \hat{s}_{\beta} + 1/2] $, where the constant term is the analogue of the zero-point energy.

For simplicity, after defining a zero-point magnetization $ M_{0\beta} = \sqrt{\hbar|\gamma|M_S/\tilde{\Lambda}_{\beta}} $ and the normalization constant $ \tilde{\Lambda}_{\beta} = 2 \mathrm{Im} \int d V \tilde{m}_{x}^{*}(\mathbf{r}) \tilde{m}_{y}(\mathbf{r}) $, the mode functions are replaced as~\cite{PhysRevB.101.125404}
\begin{equation}\label{a23}
	\left[\begin{array}{c}
	\vec{m}_{\beta}(\vec{r}) \\
	\vec{h}_{\beta}(\vec{r})
	\end{array}\right] \rightarrow {M}_{0 \beta}\left[\begin{array}{c}
	\tilde{\vec{m}}_{\beta}(\vec{r}) \\
	\tilde{\vec{h}}_{\beta}(\vec{r})
	\end{array}\right].
\end{equation}
The corresponding magnetization and magnetic field operators in the Schrödinger picture are writen as
\begin{subequations}
	\begin{eqnarray}
	\hat{\vec{m}}(\vec{r})=\sum_{\beta} {M}_{0 \beta}\left[\tilde{\vec{m}}_{\beta}(\vec{r}) \hat{s}_{\beta}+\mathrm{H.c.}\right],\label{a24a}\\
	\hat{\vec{h}}(\vec{r})=\sum_{\beta} {M}_{0 \beta}\left[\tilde{\vec{h}}_{\beta}(\vec{r}) \hat{s}_{\beta}+\mathrm{H.c.}\right].\label{a24b}
	\end{eqnarray}
\end{subequations}
As for the Kittel mode, especially, the mode function is $ \tilde{m}_K = \vec{e}_x+i\vec{e}_y $ with the coordinate vector $ \vec{e}_x $ and $ \vec{e}_y $. The zero-point magnetization is $ M_{K} = \sqrt{\hbar |\gamma|M_S/2V} $ with the saturation magnetization $ M_S $ and the volume $ V $~\cite{PhysRevB.101.125404}. Then the corresponding magnetization operator is
\begin{equation}\label{a25}
	\hat{ \vec {M}} = {M_{K}} \left( \tilde{ m}_K\hat {s}_K + \tilde{ m}_K^*\hat {s}_K^{\dagger}\right).
\end{equation}

\section{\label{app:b}Quantization of the magnetic field generated by a magnetic sphere and the interaction Hamiltonian}

In this appendix, we show the quantization of the magnetic field generated by a magnetic sphere concretely.

We start from the classical electromagnetism result that the field of a magnetic sphere with magnetization $ \vec {M} $ can be described by
\begin{equation}\label{b1}
	\vec{B}(\vec{r})=\frac{\mu_{0}}{3} \frac{R^{3}}{r^{3}}\left[\frac{3(\vec{M} \cdot \vec{r}) \vec{r}}{r^{2}}-\vec{M}\right],
\end{equation}
where $ \vec{B}(\vec{r}) $ is the magnetic field at $ \vec{r}=r\cos\theta\vec{e}_x+r\sin\theta\vec{e}_y $.
We then introduce the quantized magnetization operator from Eq.~(\ref{a25}), i.e.,
\begin{equation}\label{b2}
\hat{\vec{M}}=M_{K}\left[\left(\hat{s}_{K}+\hat{s}_{K}^{\dagger}\right) \vec{e}_{x}+i\left(\hat{s}_{K}-\hat{s}_{K}^{\dagger}\right) \vec{e}_{y}\right].
\end{equation}
Applying the above expression to Eq.~(\ref{b1}), we can obtain the operator expression of the magnetic field
\begin{eqnarray}\label{b3}
\hat{ \vec {B}}_m(\vec{r}) =&& \frac{\mu_0 R^3 M_{K}}{3r^3}\Big\{\big[ (3C_{\theta}^2-1)\hat{X}+3S_{\theta}C_{\theta}\hat{P} \big]\vec{e}_x \nonumber\\
&&+ \big[ 3S_{\theta}C_{\theta}\hat{X} + (3S_{\theta}^2-1)\hat{P} \big]\vec{e}_y \Big\},
\end{eqnarray}
where we define $ C_{\theta}=\cos\theta $, $ S_{\theta}=\sin\theta $, $ \hat{X} = \hat {s}_K + \hat {s}_K^{\dagger} $ and $ \hat{P} = i(\hat {s}_K - \hat {s}_K^{\dagger}) $ for convenience.

Then we consider the interaction between the spin and the quantized mgnetic field described by $ \hat{H}_{N-K} = (-g_e\mu_B/\hbar) \hat{ \vec {B}}\cdot \hat{\vec{S}} $ with the spin operator $ \hat{S} = \hat{S}_x \vec{e}_x + \hat{S}_y \vec{e}_y + \hat{S}_z \vec{e}_z $, which leads to the Hamiltonian
\begin{eqnarray}\label{b4}
\hat{H}_{N-K} =&& -\hbar g \big[ (3C_{\theta}^2-1)(\hat {s}_K + \hat {s}_K^{\dagger})(\hat{\sigma}^{+} + \hat{\sigma}^{-})\nonumber\\
&&+3iS_{\theta}C_{\theta}(\hat {s}_K - \hat {s}_K^{\dagger})(\hat{\sigma}^{+} + \hat{\sigma}^{-})\nonumber\\
&&-3iS_{\theta}C_{\theta}(\hat {s}_K + \hat {s}_K^{\dagger})(\hat{\sigma}^{+} - \hat{\sigma}^{-})\nonumber\\
&&+(3S_{\theta}^2-1)(\hat {s}_K - \hat {s}_K^{\dagger})(\hat{\sigma}^{+} - \hat{\sigma}^{-}) \big],
\end{eqnarray}
where the spin operators are defined as $ \hat{\sigma}^{\pm} = (\hat{S}_x \pm \hat{S}_y)/\hbar $.
Under the rotating wave approximation, the imaginary terms in the second and third lines of the above expression offset with each other, while the left parts can be simplified to the final Hamiltonian $ \hat{H}_{N-K}\simeq -\hbar g(\hat{s}_{K} \hat{\sigma}^{+}+H.c.) $. Here, the coupling strength is defined as
\begin{eqnarray}\label{b5}
g = \sqrt{\frac{|\gamma|M_s}{12\pi}}\frac{g_e \mu_0 \mu_B R^{3/2}}{(R+d)^3},
\end{eqnarray}
where the distance between the spin qubit and the surface of the magnetic sphere is $ d = r-R $ (see Fig.~\ref{fig:f1}(a)).

\providecommand{\noopsort}[1]{}\providecommand{\singleletter}[1]{#1}%

\end{document}